\RequirePackage{fix-cm}
\documentclass[smallextended]{svjour3}       
\smartqed  
\usepackage{graphicx}
\begin{document}

\title{Charged Massive Particle's Tunneling From Charged Non-Rotating Micro Black Hole
}


\author{M. J. Soleimani         \and
        N. Abbasvandi  \and Shahidan Radiman \and W.A.T. Wan Abdullah
}


\institute{M. J. Soleimani \at
              Physics Department, University of Malaya, KL, 50603, Malaysia \\
              Tel.: +603-7967 4201\\
              Fax: +603-7956 6343\\
              \email{msoleima@cern.ch}           
           \and
           N. Abbasvandi \at
              School of Applied Physics, FST, University Kebangsaan Malaysia, Bangi, 43600, Malaysia
              Tel.: +603-8921-5419
              Fax: +603-8925-6086
              \email{Niloofar@siswa.ukm.edu.my}
           \and
           Shahidan Radiman \at
           School of Applied Physics, FST, University Kebangsaan Malaysia, Bangi, 43600, Malaysia
           Tel.: +603-8921-5419
           Fax: +603-8925-6086
           \email{Shahidan@ukm.edu.my}
           \and
           W.A.T. Wan Abdullah \at
           Physics Department, University of Malaya, KL, 50603, Malaysia \\
           Tel.: +603-7967 4201\\
           Fax: +603-7956 6343\\
           \email{wat@um.edu.my}
}

\date{Received: date / Accepted: date}

\maketitle

\begin{abstract}
In the tunneling framework of Hawking radiation, charged massive particle's tunneling in charged non-rotating TeV-Scale black hole is investigated. To this end, we consider natural cutoffs as a minimal length, a minimal momentum, and a maximal momentum through a generalized uncertainty principle. We focus on the role played by these natural cutoffs on the luminosity of charged non-rotating micro black hole by taking into account the full implications of energy and charge conservation as well as the back- scattered radiation.
\keywords{Quantum Gravity \and Tunneling effect \and Micro Black Hole \and Generalized Uncertainty Principle \and Large Extra Dimensions}
\PACS{04.50.Gh \and 04.60.-m}
\end{abstract}

\section{Introduction}
\label{}
One of the most exciting consequence of models of low scale gravity \cite{arkani,antoniadis,randall1,randall2} is the possibility of production of small black holes \cite{argyres,bank,emparan,giddings} at particle colliders such as the Large Hadron Collider (LHC) as well as in Ultrahigh Energy Cosmic Ray Air Showers (UECRAS) \cite{feng,ringwald,cavaglia}. Incorporation of gravity in quantum field theory support the idea that the standard Heisenberg uncertainty principle should be reformulated by the so-called Generalized Uncertainty Principle near the Planck scale \cite{veneziano,kempf1,kempf2}. In particular, the existence of a minimum observable length is indicated by string theory \cite{amati}, TeV-Scale black hole physics \cite{meissner}, and loop quantum gravity \cite{gary}. Moreover, some black hole Gedanken experiments support the idea of existence of a minimal measurable length in a fascinating manner \cite{scardigli,adler} On the other hand, Doubly Special Relativity theories \cite{ali,das,amelino} suggest that, a test particle's momentum cannot be arbitrarily imprecise and there is an upper bound, , for momentum fluctuation. It means that there is also a maximal particle momentum. It has been shown that incorporation of quantum gravity effects in black hole physics and thermodynamics through a Generalized Uncertainty Principle (GUP) with the mentioned natural cutoffs modifies the result dramatically, specially, the final stage of black hole evaporation. Parikh and Wilczek on their pioneering work \cite{parikh} constructed a procedure to describe the Hawking radiation emitted from a Schwarzschild black hole as a tunneling through its quantum horizon. The emission rate (tunneling probability) which arising from the reduction of the black hole mass, is related to the change of black hole entropies before and after the emission. In this article, charged particle's tunneling from charged non-rotating micro black hole is investigated. We consider a more general framework of GUP that admits a minimal length, minimal momentum, and maximal momentum to study the effects of natural cutoffs on the tunneling mechanism and luminosity of charged non-rotating TeV-Scale black holes with extra dimensions in Arkani-Hamed, Dimopoulos and Dvali (ADD) model \cite{arkani} in the context of this GUP. The calculation shows that, the emission rate satisfies the first law of black hole thermodynamics. The paper is organized as follows: In section 2, we introduce a generalized uncertainty principle with minimal length, minimal momentum, and maximal momentum. In section 3, we obtain an expression for emission rate of charged particle from charged non-rotating micro black hole based on the ADD model and the mentioned GUP. We consider the back scattering of the emitted radiation taking into account energy and charge conservation to evaluate the luminosity of TeV-Scale black hole in presence of natural cutoffs. The last part is the discussion and calculation. 
\section{Generalized uncertainty principle}
\label{}
The existence of a minimal measurable length of the order of the Planck length, ${l_p} \sim {10^{ - 35}}m$ was indicated by most of quantum gravity approaches \cite{maggiore1,maggiore2} which modifies the Heisenberg Uncertainty principle (HUP) to the so-called generalized (Gravitational) Uncertainty Principle (GUP). The minimal position uncertainty,$\Delta {x_0}$ , could be not made arbitrarily small toward zero \cite{kempf1} in the GUP framework due to its essential restriction on the measurement precision of the particle's position. On the other hand, Doubly Special Relativity (DSR) theories \cite{amelino} has been considered that existence of a minimal measurable length would restrict a test particle's momentum to take arbitrary values and therefore there is an upper bound for momentum fluctuation \cite{magueijo1,corte}. So, there is a maximal particle's momentum due to the fundamental structure of spacetime at the Planck scale \cite{magueijo2,magueijo3}. Based on the above arguments, the GUP that predicts both a minimal length and a maximal momentum can be written as follows \cite{ali,das}:
\begin{equation} \label{1}
\Delta x\Delta p \ge \frac{\hbar }{2}\left[ {1 - \alpha \langle \Delta p\rangle  + 2{\alpha ^2}{{\langle \Delta p\rangle }^2}}  \right]
\end{equation}
The relation (1) can lead us to the following commutator relation:
\begin{equation} \label{2}
[x,p] = i\hbar \left( {1 - \alpha p + 2{\alpha ^2}{p^2}} \right)\
\end{equation}
Where $\alpha$ is GUP dimensionless positive constant of both minimal length and maximal momentum that depends on the details of the quantum gravity hypothesis. It has developed that particle's momentum cannot be zero if the curvature of spacetime becomes important and uts effects are taken into account \cite{hinrichsen,zarei}. In fact, there appears a limit to the precision which the corresponding momentum can be expressed as a nonzero minimal uncertainty in momentum measurement. Based on this more general framework as a consequence of small correction to the canonical commutation relation, this GUP can be represented as \cite{pedram}
\begin{equation} \label{3}
\Delta x\Delta p \ge \hbar \left( {1 - \alpha {l_p}\Delta p + {\alpha ^2}{l_p}^2{{(\Delta p)}^2} + {\beta ^2}{l_p}^2{{(\Delta x)}^2}} \right)\
\end{equation} 
which in extra dimensions can be written as follows \cite{abbasvandi}
\begin{equation} \label{4}
\Delta {x_i}\Delta {p_i} \ge \hbar \left( {1 - \alpha {l_p}(\Delta {p_i}) + {\alpha ^2}{l_p}^2{{(\Delta {p_i})}^2} + {\beta ^2}{l_p}^2{{(\Delta {x_i})}^2}} \right)\
\end{equation}
Here, $\alpha$ and $\beta$ are dimensionless positive coefficients which are independent of $\Delta x$ and $\Delta p$. In general they may depend on expectation value of x and p. according to the generalized Heisenberg algebra, we suppose that operators of position and momentum obey the following commutation relation:
\begin{equation} \label{5}
[x,p] = i\hbar \left( {1 - \alpha p + {\alpha ^2}{p^2} + {\beta ^2}{x^2}} \right)\
\end{equation}
In what follows, we use this more general framework of GUP to find the tunneling rate of emitted particles through charged non-rotating TeV-Scale black holes. 

\section{Tunneling Mechanism}
The idea of large extra dimensions might allow studying interactions at Trans-Planckian energies in particle colliders and the ADD model used d new large space-like dimensions. So, in order to investigate the Hawking radiation via tunneling from charged non- rotating TeV-Scale black holes of higher dimensional, a natural candidate is that of Reissner-Nordstrom d-dimensional modified solution in presence of generalized uncertainty principle \cite{abbasvandi}. In this case, the line element of d-dimensional Reissner-Nordstrom solution of Einstein field equation is given by \cite{myers} 
\begin{equation} \label{6}
d{s^2} = f(r){c^2}d{t^2} - {f^{ - 1}}(r)d{r^2} - {r^2}d\Omega _{d - 2}^2 = {g_{\mu \nu }}d{x^\mu }d{x^\nu }\
\end{equation}
where ${\Omega _{d - 2}}$ is the metric of the unit ${S^{d - 2}}$ as ${\Omega _{d - 2}} = \frac{{2{\pi ^{\frac{{d - 1}}{2}}}}}{{\Gamma (\frac{{d - 1}}{2})}}$, and
\begin{equation} \label{7}
\begin{array}{l}
f = f(M,Q,r) = 1 - \frac{{{\omega _{d - 2}}M}}{{{r^{d - 3}}}} + \frac{{{\omega _{d - 2}}{Q^2}}}{{2(d - 3){\Omega _{d - 2}}{r^{2(d - 3)}}}}\\
\end{array}
\end{equation}
where ${\omega _{d - 2}} = \frac{{16\pi }}{{(d - 2){\Omega _{d - 2}}}}$. Here $M$ and $Q$ are the mass and electric charge of the black hole, respectively, units ${G_d} = c = \hbar  = 1$ are adopted throughout this manuscript. The black hole has an outer/inner horizon located at 
\begin{equation} \label{8}
r_ \pm ^{d - 3} = \frac{{{\omega _{d - 2}}}}{2}\left[ {M \pm \sqrt {{M^2} - \frac{{(d - 2){Q^2}}}{{8\pi (d - 3)}}} } \right]
\end{equation}
Therefore, the event horizon shrinks, and the inner one appears, when the black hole becomes charged the inner radius is related to the amount of charge and the outer one ${r_ + }$ corresponds to the radius of Schwarzschild black hole. In this case, equation (8) can be rewritten as follows
\begin{equation} \label{9}
{r_ + } = {\left( {\frac{{{\omega _{d - 2}}}}{2}\left[ {M + \sqrt {{M^2} - \frac{{(d - 2){Q^2}}}{{8\pi (d - 3)}}} } \right]} \right)^{\frac{1}{{d - 3}}}}
\end{equation}
In order to apply the semi- classical tunneling analysis, one can find a proper coordinate system for the black hole metric where all the constant lines are flat and the tunneling path is free of singularities. In this manner, Painleve' coordinates are suitable choices. In these coordinates, the d-dimensional Reissner-Nordstrom metric is given by 
\begin{equation} \label{10}
d{s^2} =  - fd{t^2} \pm 2\sqrt {1 - f} dtdr + d{r^2} + {r^2}d\Omega _{d - 2}^2
\end{equation}
which is stationary, non-static, and non-singular at the horizon and plus (minus) sign corresponds to the space-time line element of the outgoing (incoming) particles across the event horizon, respectively. 
The trajectory of charged massive particles as a sort of de Broglie s-wave can be approximately determined as \cite{jiang,zhang}
\begin{equation} \label{11}
\dot r = \frac{{dr}}{{dt}} =  - \frac{{{g_{tt}}}}{{2{g_{tr}}}} =  \pm \frac{f}{{2\sqrt {1 - f} }}
\end{equation}
where the plus (minus) sign denotes the radial geodesics of the outgoing (incoming) charged particles tunneling across the event horizon, respectively. We incorporate quantum gravity effects in the presence of the minimal length, minimal momentum, and maximal momentum via the GUP which motivates modification of the standard dispersion relation in the presence of extra dimensions based on ADD model. 
If the GUP is a fundamental outcome of quantum gravity proposal, it should appear that the de Broglie relation is as follows \cite{saghafi}  
\begin{equation} \label{12}
{\lambda _ \pm } = \frac{{{p_i}}}{{2{\beta ^2}l_p^2}}\left( {1 \pm \sqrt {1 - \frac{{4{\beta ^2}l_p^2(1 - \alpha {l_p}{p_i} + {\alpha ^2}l_p^2p_i^2}}{{p_i^2}}} } \right)
\end{equation}
One can find easily that positive sign does not recover ordinary relation in the limit $\alpha  \to 0$ and $\beta  \to 0$. So, we consider the minus sign as 
\begin{equation} \label{13}
{\lambda _ - } = \frac{1}{{{p_i}}}\left( {1 - \alpha {l_p}{p_i} + {\alpha ^2}l_p^2p_i^2} \right)\left( {1 + \frac{{{\beta ^2}l_p^2}}{{p_i^2}}(1 - \alpha {l_p}{p_i} + {\alpha ^2}l_p^2p_i^2)} \right)
\end{equation}
Or equivalently 
\begin{equation} \label{14}
\varepsilon  = E\left( {1 - \alpha {l_p}E + {\alpha ^2}l_p^2{E^2}} \right)\left( {1 + \frac{{{\beta ^2}l_p^2}}{{{E^2}}}(1 - \alpha {l_p}E + {\alpha ^2}l_p^2{E^2})} \right)
\end{equation}
Here, for investigating Hawking radiation of charged massive particles from the event horizon of charged non-rotating micro black hole, we use this more general uncertainty principle and take into consideration the response of background geometry to radiated quantum of energy E with GUP correction, i.e. $\varepsilon$. The emitted particle which can be treated as a shell of energy $\varepsilon$ and charge q, moves on the geodesics of a space-time with central mass $M - \varepsilon$  substituted for M and charge parameter $Q - q$ replaced with $Q$. We set the total Arnowitt-Deser-Misner (ADM) mass, M, and the ADM charge of the space-time to be fixed but allow the hole mass and charge to fluctuate and replace M by $M - \varepsilon$ and $Q$ by $Q-q$ both in the metric and the geodesic equation. So, the outgoing radial geodesics of the charged massive particle tunneling out from the event horizon and the non-zero component of electromagnetic potential are, 
\begin{equation} \label{15}
\dot r = \frac{{f(M - \varepsilon ,Q - q,r)}}{{2\sqrt {1 - f(M - \varepsilon ,Q - q,r)} }}
\end{equation}
and,
\begin{equation} \label{16}
{A_t} = \frac{{Q - q}}{{(d - 3){\Omega _{d - 2}}{r^{d - 3}}}}
\end{equation}
So, the Lagrangian for the matter-gravity system is 
\begin{equation} \label{17}
L = {L_m} + {L_e}
\end{equation}
where ${L_e} =  - \frac{1}{4}{F_{\mu \nu }}{F^{\mu \nu }}$ is the Lagrangian function of the electromagnetic field corresponding to the generalized coordinates ${A_\mu } = ({A_t},0,0,0)$ \cite{makela}.
\\We assume the tunneling mechanism as a semi-classical method producing Hawking radiation. In this case, using WKB approximation, the emission rate of tunneling massive charged particle can be obtained from the imaginary part of the particle action at the stationary phase for the tunneling trajectory, namely \cite{srinivasan,parikh2}
\begin{equation} \label{18}
\Gamma  \sim \exp \left( { - 2{\mathop{\rm Im}\nolimits} I} \right)
\end{equation}
Assuming the generalized coordinate ${A_t}$ is an ignorable one, to eliminate this degree of freedom completely, we can obtain the action of the matter-gravity system as 
\begin{equation} \label{19}
I = \int\limits_{{t_i}}^{{t_f}} {\left( {L - {P_{{A_t}}}{{\dot A}_t}} \right)dt}  = \int\limits_{{r_i}}^{{r_f}} {\left[ {\int\limits_{\left( {0,0} \right)}^{\left( {{p_r},{p_{{A_t}}}} \right)} {\left( {\dot rd{{p'}_r} - {{\dot A}_t}d{{p'}_{{A_t}}}} \right)} } \right]} \frac{{dr}}{{\dot r}}
\end{equation}
where ${r_i}$ and ${r_f}$ are the location of the event horizon corresponding ${t_i}$ and ${t_f}$, respectively before and after the particle of energy $\varepsilon$ and charge $q$ tunnels out, in which ${p_{{A_t}}}$, and ${p_r}$ are the canonical momentum conjugate to the coordinates ${A_t}$, and r, respectively. 
\\In order to consider the effect of quantum gravity, the commutation relation between the radial coordinate components and conjugate momentums should be modified based on Eq. (1) $\&$ (2) of the expressed GUP as follows \cite{saghafi}
\begin{equation} \label{20}
\left[ {r,{p_r}} \right] = i\left( {1 - \alpha {l_p}{p_r} + {\alpha ^2}l_p^2p_r^2} \right)
\end{equation} 
So, as it is clear from the more general GUP and based on Eq. (5) the commutation relation should be modified as
\begin{equation} \label{21}
\left[ {r,{p_r}} \right] = i\left( {1 - \alpha {l_p}{p_r} + {\alpha ^2}l_p^2p_r^2 + {\beta ^2}l_p^2{r^2}} \right)
\end{equation}
In the classical limit it is replaced by Poisson bracket as follows
\begin{equation} \label{22}
\left\{ {r,{p_r}} \right\} = \left( {1 - \alpha {l_p}{p_r} + {\alpha ^2}l_p^2p_r^2 + {\beta ^2}l_p^2{r^2}} \right)
\end{equation}
Now, we apply the deformed Hamiltonian equation, 
\begin{equation} \label{23}
\dot r = \left\{ {r,H} \right\} = \left\{ {r,{p_r}} \right\}{\left. {\frac{{dH}}{{dr}}} \right|_r},{\left. {dH} \right|_{(r,{A_t},{p_t})}} = d(M - \varepsilon )
\end{equation}
and
\begin{equation} \label{24}
{\dot A_t} = \frac{{dH}}{{d{p_{{A_t}}}}} = \frac{{d{{E'}_Q}}}{{d{p_{{A_t}}}}},{\left. {dH} \right|_{({A_t},r,{p_r})}} = {A_t}d(Q - q)
\end{equation}
Into Eq. (19) as the Hamiltonian is $H = M - \varepsilon '$, one can ${p^2} \simeq {\varepsilon '^2}$, $p \simeq \varepsilon '$ and eliminate the momentum in the favor of the energy in integral (19) and switching the order of integration yield the imaginary part of the action as follows 
\begin{equation} \label{25}
\begin{array}{l}
{\mathop{\rm Im}\nolimits} I = {\mathop{\rm Im}\nolimits} \int\limits_{{r_i}}^{{r_f}} {\int\limits_{(M,Q)}^{(M - \varepsilon ,Q - q)} {\left[ {(1 - \alpha {l_p}\varepsilon ' + {\alpha ^2}l_p^2{{\varepsilon '}^2} + \beta l_p^2{r^2})d(M - \varepsilon ') - \frac{{Q - q'}}{{(d - 3){\Omega _{d - 2}}{r^{d - 3}}}}d(Q - q')} \right]} } \frac{{dr}}{{\dot r}}\\
= {\mathop{\rm Im}\nolimits} \int\limits_{{r_i}}^{{r_f}} {\int\limits_{(M,Q)}^{(M - \varepsilon ,Q - q)} {\frac{{2\sqrt {1 - f(M - \varepsilon ',Q - q',r)} }}{{f(M - \varepsilon ',Q - q',r)}}} } \left[ {(1 - \alpha {l_p}\varepsilon ' + {\alpha ^2}l_p^2{{\varepsilon '}^2} + \beta l_p^2{r^2})d(M - \varepsilon ')- \frac{{Q - q'}}{{(d - 3){\Omega _{d - 2}}{r^{d - 3}}}}d(Q - q')} \right]dr
\end{array}
\end{equation} 
The r integral can be evaluated by deforming the contour of the single pole at the outer horizon. During the r integral first, we find 
\begin{equation} \label{26}
\begin{array}{l}
{\mathop{\rm Im}\nolimits} I = {\mathop{\rm Im}\nolimits} \int\limits_0^\varepsilon  {2( - \pi i){r_ + }\left( {M - \varepsilon ',Q - q} \right)\left( {1 - \alpha {l_p}\varepsilon ' + {\alpha ^2}l_p^2{{\varepsilon '}^2} + \beta l_p^2} \right)} d( - \varepsilon ')\\
- \int\limits_0^{Q - q} {2( - \pi i){r_ + }} \left( {M - \varepsilon ,Q - q'} \right)\frac{{(Q - q')}}{{(d - 3)\pi {r^{d - 3}}}}d(Q - q')
\end{array}
\end{equation}
This allows us to consider the leading order correction to be just proportional to second order of $\alpha {l_p}$ and also second order of $\beta {l_p}$ for simplicity without loss of generality. In this regards, we can finish the integration by applying Taylor series and obtain the imaginary part of the action.
Although, the integral (26) is complicated, one can find such terms as an example for $d = 5$ as follows
\begin{equation} \label{27}
\begin{array}{l}
{\mathop{\rm Im}\nolimits} {I_{d = 5}} \approx ... + 4096\alpha {l_p}\pi {M^2} - 8192\alpha {l_p}\pi ME\left( {1 - \alpha {l_p}E + {\alpha ^2}l_p^2{E^2}} \right)\left( {1 + \frac{{{\beta ^2}l_p^2}}{{{E^2}}}(1 - \alpha {l_p}E + {\alpha ^2}l_p^2{E^2})} \right)\\
+ 4096\alpha {l_p}\pi {E^2}{\left( {1 - \alpha {l_p}E + {\alpha ^2}l_p^2{E^2}} \right)^2}{\left( {1 + \frac{{{\beta ^2}l_p^2}}{{{E^2}}}(1 - \alpha {l_p}E + {\alpha ^2}l_p^2{E^2})} \right)^2} - 192\alpha {l_p}{Q^2} + ...
\end{array}
\end{equation}
Substituting (26) into (18), the tunneling probability of charged particles from charged non-rotating TeV-Scale black holes is obtained as 
\begin{equation} \label{28}
\begin{array}{l}
\Gamma  = \exp \left( { - 2{\mathop{\rm Im}\nolimits} I} \right)\\
 \simeq \exp \left[ {{\mathop{\rm Im}\nolimits} \int\limits_0^\varepsilon  {2( - \pi i){r_ + }(M - \varepsilon ',Q - q)(1 - \alpha {l_p}} \varepsilon ' + {\alpha ^2}l_p^2{{\varepsilon '}^2} + \beta l_p^2)d( - \varepsilon ')- \int\limits_0^{Q - q} {2( - \pi i){r_ + }} (M - \varepsilon ,Q - q')\frac{{(Q - q')}}{{(d - 3)\pi {r^{d - 3}}}}d(Q - q')} \right]\\
 = \exp \left( {\Delta s} \right)
\end{array}
\end{equation}
where $\Delta s$ is the difference in black hole entropies before and after emission \cite{kraus,vagenas1,vagenas2,vagenas3,medved,vagenas4}. It was shown that the emission rates on the high energy scales corresponds to differences between the counting of states in the micro canonical and in the canonical ensembles \cite{parikh3,akhmedov}. By performing integration on (26), one can find that the first order of E in the exponential gives a thermal, Boltzmannian spectrum. The existence of extra terms in relation (27) shows that the radiation is not completely thermal. In fact, these extra terms enhance the non-thermal character of the radiation. Also, it is easy to find that  should be greater than  in any stage of tunneling process. This tunneling rate compared to the tunneling rate which is calculated in \cite{wu}, obviously shows that by considering all natural cutoffs in generalized uncertainty principle relation, many additional terms are appeared. The additional terms shows strong deviation of micro black holes radiation from ordinary thermal radiation. 
\section{Back-scattering and luminosity}
It has shown \cite{hawking} that black holes radiate a thermal spectrum of particles. So, micro black holes emit black body radiation at the Hawking temperature.
Following a heuristic argument \cite{adler1}, the energy of the Hawking particles is $\Delta E \approx c\Delta p$ and it is deduced for the Hawking temperature of black hole based on LED scenario,
\begin{equation} \label{29}
{T_H} \simeq \frac{{(d - 3)\Delta p}}{{4\pi }}
\end{equation}
which $\frac{{(d - 3)}}{{4\pi }}$ is a calibration factor in d-dimensional space-time. By saturating inequality (4), one can find momentum uncertainty in terms of position uncertainty as follows 
\begin{equation} \label{30}
\Delta {P_i} = \left( {\frac{{\alpha {l_p} + \Delta {x_i}}}{{4{\alpha ^2}l_p^2}}} \right)\left( {1 \pm \sqrt {\frac{{4{\alpha ^2}l_p^2(1 + {\beta ^2}l_p^2{{(\Delta {x_i})}^2}}}{{{{(\alpha {l_p} + \Delta {x_i})}^2}}}} } \right)
\end{equation}
So, the modified black hole Hawking temperature in the presence of natural cutoffs becomes 
\begin{equation} \label{31}
{T_H} = \frac{\begin{array}{l}
	\\
	(d - 3)(2{r_ + } + \alpha {l_p})
	\end{array}}{{16\pi {\alpha ^2}l_p^2}}\left( {1 - \sqrt {1 - \frac{{4{\alpha ^2}l_p^2(1 + {\beta ^2}l_p^2{r_ + }^2)}}{{{{(2{r_ + } + \alpha {l_p})}^2}}}} } \right)
\end{equation}
Based on equation (31), GUP give rise to the existence of a minimal mass of a charged non-rotating micro black hole given by 
\begin{equation} \label{32} 
M_{\min }^{GUP} = \frac{{(d - 2){\Omega _{d - 2}}{l_p}}}{{16\pi {G_d}}}\left[ \begin{array}{l}
{\left[ {\frac{{ - \alpha {l_p} + ({\alpha ^2}l_p^2 + 3{\alpha ^2}l_p^2{{(1 - {\alpha ^2}{\beta ^2}l_p^4)}^{\frac{1}{2}}}}}{{2(1 - {\alpha ^2}{\beta ^2}l_p^2)}}} \right]^{\frac{1}{{d - 3}}}}\\
+ \frac{{{Q^2}8\pi {G_d}}}{{(d - 2)(d - 3){{\left[ {\frac{{ - \alpha {l_p} + ({\alpha ^2}l_p^2 + 3{\alpha ^2}l_p^2{{(1 - {\alpha ^2}{\beta ^2}l_p^4)}^{\frac{1}{2}}}}}{{2(1 - {\alpha ^2}{\beta ^2}l_p^2)}}} \right]}^{\frac{1}{{d - 3}}}}}}
\end{array} \right]{M_p}
\end{equation}
\begin{figure} [h]
	\center
	\includegraphics*[width=0.5\textwidth]{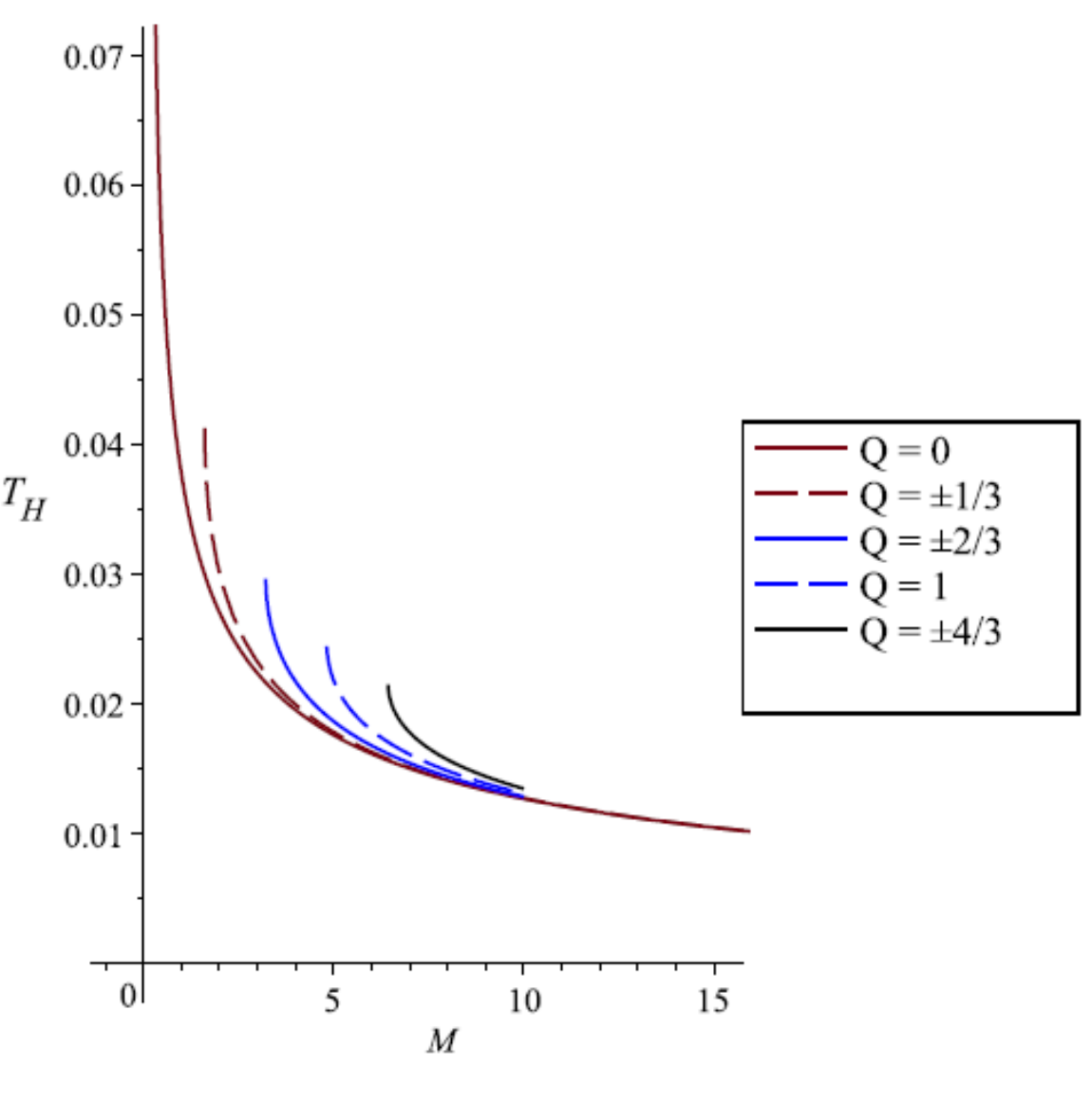}
	\caption{Hawking Temperature for different amount of Charges in presence of GUP}
	\label{fig2}
\end{figure}
So, there are some black hole remnants without radiation based on (32).
A radiated particle state corresponding to an arbitrary finite number of virtual pairs inside the black hole event horizon is as follows \cite{modak} 
\begin{equation} \label{33}
\left| \psi  \right\rangle  = N\sum {e^{ - \frac{{\pi n\varepsilon }}{{\hbar k}}}}\left| {n_{out}^{(L)}} \right\rangle  \otimes \left| {n_{out}^{(R)}} \right\rangle 
\end{equation}
where ${N^2} = \frac{{{e^{\gamma \varepsilon }}}}{{{e^{\gamma \varepsilon }} - 1}}$
is a normalization constant and $\kappa $ is the surface gravity. This quantum state is transformed with respect to an observer outside the horizon. In order to obtain the average particle number in the energy state $\varepsilon $ with respect to an observer, one can trace out the inside degrees of freedom to yield the reduced density matrix of the form 
\begin{equation} \label{34}
{\rho _{reduced}} = \left( {1 - \exp ( - \frac{{2\pi \varepsilon }}{{\hbar k}})} \right)\sum\limits_{n = 0}^\infty  {{e^{ - \gamma n\varepsilon }}} \left| {n_{out}^{(R)}} \right\rangle  \otimes \left\langle {n_{out}^{(R)}} \right|
\end{equation}
In this regards, the number distribution with respect to $\varepsilon $ is given by
\begin{equation} \label{35}
\left\langle {{n_\varepsilon }} \right\rangle  = trace\left( {n{\rho _{reduced}}} \right) = \frac{1}{{{e^{\gamma \varepsilon }} - 1}}
\end{equation}
where $\gamma  = \frac{1}{{{T_H}}}$. Whenever a particle is radiated from the micro black hole event horizon, its wave function satisfies a wave equation with an effective potential that depends on outer event horizon. As the potential represents a barrier to the outgoing radiation, so one part of the radiation is back-scattered. 
\\In this way, it can be shown that the distribution $\left\langle {{n_\varepsilon }} \right\rangle $ for the Hawking radiation will be modulated by grey body factor \cite{torres} which for a charged non-rotating TeV-Scale black hole is given as  
\begin{equation} \label{36}
\Lambda  = 4{\varepsilon ^2}r_ + ^2
\end{equation}
which $\Lambda $   is the standard approximated grey body factor. In this way, one can take energy and charge conservation into account \cite{torres} and get the straightforward result by substituting Eq. (14) into (36). So, we obtain
\begin{equation} \label{37}
{\Lambda _{EC}} = 4{\left[ {E\left( {1 - \alpha {l_p}E + {\alpha ^2}l_p^2{E^2}} \right)\left( {1 + \frac{{{\beta ^2}l_p^2}}{{{E^2}}}(1 - \alpha {l_p}E + {\alpha ^2}l_p^2{E^2})} \right)} \right]^2}r_ + ^2(M - \varepsilon ,Q - q)
\end{equation}
\begin{figure} [h]
	\center
	\includegraphics[width=0.5\textwidth]{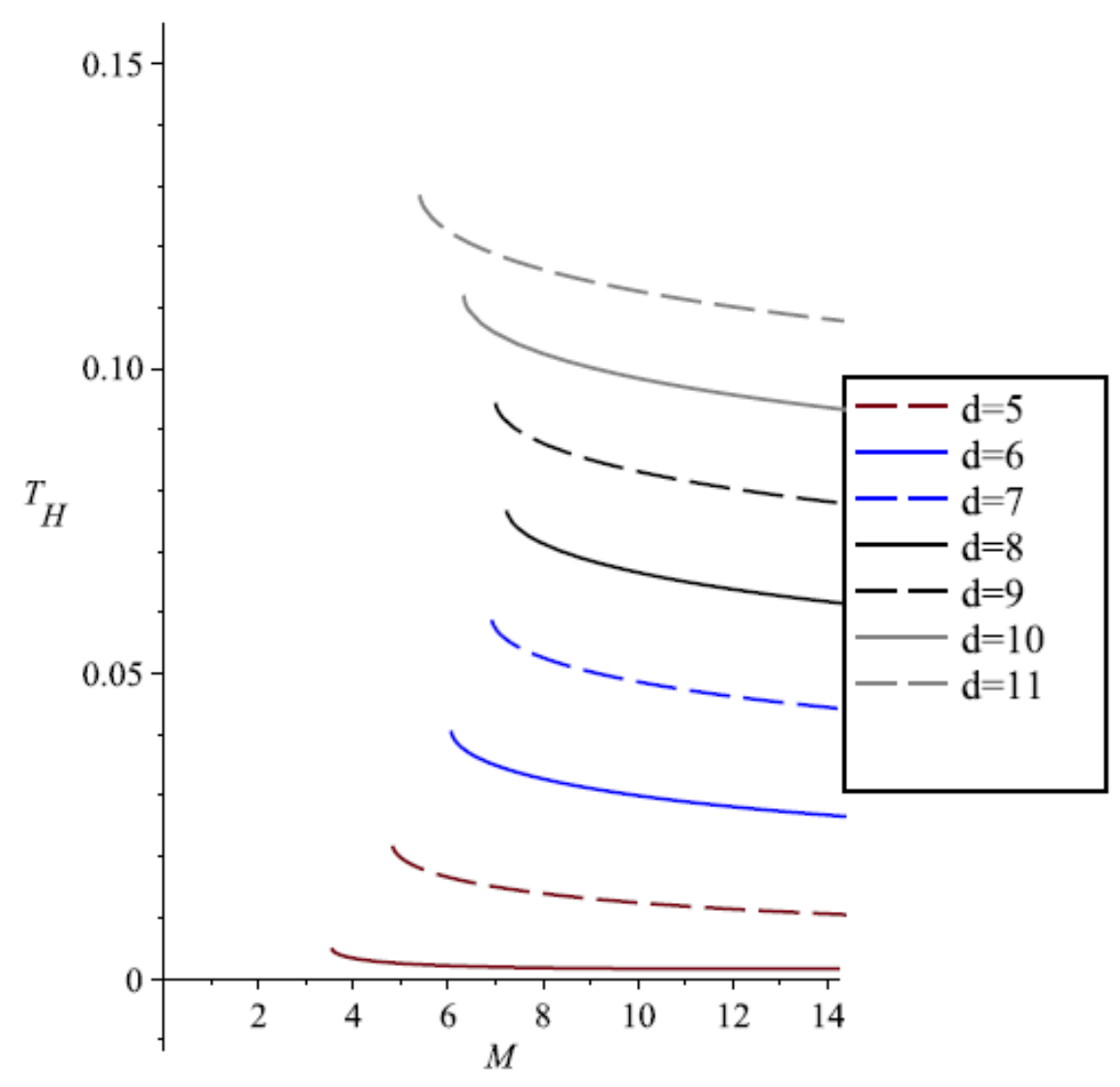}
	\caption{Hawking Temperature with respect to the mass in terms of GUP}
	\label{fig1}
\end{figure}
On the other hand, if we consider the full consequences of energy and charge conservation,for total flux, including back-scattering \cite{fabbri}, the luminosity modulated according to the gray body factor, has to be written as 
\begin{equation} \label{38}
\begin{array}{l}
{L^d}(M) = \frac{1}{{2\pi }}\int_0^{M - M_{\min }^{GUP}} {\left\langle {{n_\varepsilon }} \right\rangle } {\Lambda _{EC}}\varepsilon d\varepsilon  = \\
= \frac{1}{{2\pi }}\int_0^{M - M_{\min }^{GUP}} {\frac{{4{{\left[ {E\left( {1 - \alpha {l_p}E + {\alpha ^2}l_p^2{E^2}} \right)\left( {1 + \frac{{{\beta ^2}l_p^2}}{{{E^2}}}(1 - \alpha {l_p}E + {\alpha ^2}l_p^2{E^2})} \right)} \right]}^2}r_ + ^2(M - \varepsilon ,Q - q)d\varepsilon }}{{\exp \left( {\frac{{16\pi {\alpha ^2}l_p^2\varepsilon }}{{(d - 3)(2{r_ + } + \alpha {l_p})}}{{\left[ {1 - \sqrt {1 - \frac{{4{\alpha ^2}l_p^2(1 + l_p^2{\beta ^2}r_ + ^2)}}{{{{(2{r_ + } + \alpha {l_p})}^2}}}} } \right]}^{ - 1}}} \right) - 1}}} 
\end{array}
\end{equation}
It is important to remark that the total luminosities for micro black hole would be ten times bigger if we neglect back scattering effect. We are taking into account in the integration limits that the maximum energy of a radiated particle could be $M - M_{\min }^{GUP}$. Equation (38) gives larger luminosities for smaller masses. The results show that in large extra dimension scenario, Hawking temperature of charged black hole  increases and leads to faster decay and less classical behaviors for black holes (Figure\ref{fig1}). On the other hand, it has shown \cite{gingrich,landsberg} that the allowed particles forming the black hole at the LHC are quarks, antiquarks, and gluons which formed nine possible electric charge states: $\pm \frac{4}{3}, \pm 1, \pm \frac{2}{3}, \pm \frac{1}{3},0$. In this case, as far as the electric charge of the black hole increases, the minimum mass and its order of magnitude increase and the temperature peak displaces to the lower temperature (see Figure\ref{fig2}). As the Eq. (38) is related to the black hole temperature, based on the above arguments, the luminosity of charged non-rotating TeV-Scale black hole has different amount with respect to the charge of black hole and also extra dimensions.
\section{Conclusion and Discussion}
In this paper, we have investigated Hawking radiation of the charged massive particles as a semi-classical tunneling process from the charged non-rotating micro black hole. In this respect, we considered possible effect of natural cutoffs as a minimal length, a maximal momentum, and a minimal momentum on the tunneling rate. We have shown that in the presence of Generalized Uncertainty Principle, the tunneling rate of charged massive particle is deviated from thermal emission. In order to study the evolution of the TeV-Scale micro black hole as it evaporates respecting energy and charge conservation, we have also modified the grey-body factor, which allows considering the effect of the back scattered emitted radiation. We have calculated Hawking temperature based on the GUP admitted a minimal length, a maximal momentum, and a minimal momentum. The adopted GUP predict a minimal mass remnant with respect to the charge of black hole. So, we have been able to derive an expression for the luminosity that takes into account natural cutoffs in presence of large extra dimension based on ADD scenario for different amount of charge of black hole (Figure\ref{fig1} and Figure\ref{fig2}). The investigation implies that, considering natural cutoffs in the presence of LED, information conservation of charged non-rotating micro black hole is still possible. 

\bibliographystyle{elsarticle-num} 

\begin{thebibliography}{}
\bibitem{arkani}Arkani-Hame, N., Dimoulos, S., Dvali, G., Phys. Lett. B, 429 263 (1998).  
\bibitem{antoniadis}Antoniadis, I., Arkani-Hamed, N., Dimopoulos, S., Dvali, G., Phys. Lett. B, 436 257 (1998). 
\bibitem{randall1}Randall, L., Sundrum, R., Phys. Rev. Lett., 83 3379 (1999). 
\bibitem{randall2}Randall, L., Sundrum, R., Phys. Rev. Lett., 83 4690 (1999). 
\bibitem{argyres}Argyres, P. C., Dimouplos, S., March-Russell, J., Phys. Lett. B, 441 96 (1998).  
\bibitem{bank}Bank, T., Fischler, W., preprint hep-th/9906038
\bibitem{emparan}Emparan, W., Horowitz, G. T., Myers, R. C., Phys. Rev. Lett., 85 499 (2000). 
\bibitem{giddings}Giddings, S. B., Thomas, S., Phys. Rev. D, 77 045027 (2008). 
\bibitem{feng}Feng, J. L., Shapere, Phys. Rev. Lett., 88 135 (2002). 
\bibitem{ringwald}Ringwald, A., Tu, H., Phys. Lett. B, 525 135 (2002). 
\bibitem{cavaglia}Cavagli'a, M., Int. J. Mod. Phys. A, 18 1843 (2003).  
\bibitem{veneziano}Veneziano, G., Eur. Phys. Lett., Vol.2 No.3 p.199 (1986). 
\bibitem{kempf1}Kempf, A., Mangno, G., Mann, R. B., Phys. Rev. D Vol.52 No.2 pp. 1108-1118 (1995). 
\bibitem{kempf2}Kempf, A., Mango, G., Phys. Rev. D, Vol.55 No.12 pp. 7909-7920 (1997). 
\bibitem{amati}Amati, D., Phys. Lett. B, Vol.216 No.1-2 pp. 41-47 (1989). 
\bibitem{meissner}Meissner, K. A., Classical and Quantum Gravity, Vol.21 No.22 pp. 5245-5251 (2004). 
\bibitem{gary}Gary, L. J., IJMPA, Vol.10 No.2 p. 145 (1995). 
\bibitem{scardigli}Scardigli, F., Phys. Let. B, Vol.452 pp. 39-44 (1999). 
\bibitem{adler}Adler, R. J., American Journal of Physics, Vol.78 No.9 pp. 925 (2010).  
\bibitem{ali}Ali, A. F., Das, S., Vagenas, E. C., Phys. Rev. D, Vol.48 No.4 Article ID 044013 (2011). 
\bibitem{das}Das, S., Vagenas, E. C., Phys. Rev. lett.Vol.101 Article ID 221301 (2008) .  
\bibitem{amelino}Amelino-Camelia, G., Int. J. Mod. Phys. D, 11 (2000) 35; Amelino-Camelia, G., Nature, 418 34 (2002); Amelino-Camelia, G., Int. J. Mod. Phys. D, 11 1643 (2002); Kowalski-Glikman, J., Lect. Notes phys., 669 131 (2005). 
\bibitem{parikh}Parikh, M. K., Wilczek, F., Phys. Rev. Lett., 85 5042 (2000). 
\bibitem{maggiore1}Maggiore, M., Phys. Lett. B, 304 65 (1993). 
\bibitem{maggiore2}Maggiore, M., Phys. Lett. B, 319 83 (1993). 
\bibitem{magueijo1}Magueijo, J., Smolin, L., Phys. Rev. D, Vol.67 Article ID 044017 (2003). 
\bibitem{corte}Corte's, J. L., Gamba, J., Phys. Rev. D, Vol.71 No.6 Article ID 065015 (2005).  
\bibitem{magueijo2}Magueijo, J., Smolin, L., Phys. Rev. Lett., 88 190403 (2002). 
\bibitem{magueijo3}Magueijo, J., Smolin, L., Phys. Rev. D, 71 026010 (2005). 
\bibitem{hinrichsen}Hinrichsen, H., Kempf, A., J. Math. Phys., 37 2121 (1996).  
\bibitem{zarei}Zarei, M., Mirza, B., Phys. Rev. D, Vol.79 No.12 Article ID 125007 (2009).  
\bibitem{pedram}Pedram, P., Nozari, K., Taheri, S. H., JHEP, 03 093 (2011). 
\bibitem{abbasvandi}Abbasvandi, N., Soleimani, M. J., Wan Abdullah, W.A.T., Shahidan, Radiman, submitted to IJMPA.
\bibitem{myers}Myers, R. C., Perry, M. J., Ann. Of Phys., 172 304-347 (1986). 
\bibitem{jiang}Jiang, Q. Q., Wu, S. Q., Phys. Lett. B, 635 151 (2006). 
\bibitem{zhang}Zhang, J. Y., Zhao, Z., JHEP, 10 055 (2005). 
\bibitem{saghafi}Nozari, k., Saghafi, S., JHEP, 11 005 (2012).
\bibitem{makela}Makela, J., Pepo, P., Phys. Rev. D, 57 4899 (1998). 
\bibitem{srinivasan}Srinivasan, K., Padmanban, T., Phys. Rev. D, 60 024007 (1999). 
\bibitem{parikh2}Parikh, M. K., Int. J. Mod. Phys. D, 13 2355 (2004).  
\bibitem{kraus}Kraus, P., Wilczek, F., Nucl. Phys. B, 433 403 (1995). 
\bibitem{vagenas1}Vagenas, E. C., Phys. Lett. B, 503 399 (2001).  
\bibitem{vagenas2}Vagenas, E. C., Mod. Phys. Lett. A, 17 609 (2002). 
\bibitem{vagenas3}Vagenas, E. C., Phys. Lett. B, 533 302 (2002). 
\bibitem{medved}Medved, A. J. M., Class. Quant. Grav., 19 589 (2002). 
\bibitem{vagenas4}Vagenas, E. C., Phys. Lett. B, 559 65 (2003). 
\bibitem{parikh3}Parikh, M. K.,  Int. J. Mod. Phys. D, 13 2351 (2004).  
\bibitem{akhmedov}Akhmedov, E. T., Akhmedova, V., Singleton, D., Phys. Lett. B, 642 124 (2006).  
\bibitem{wu}Wu, Shuang-Qing., et al., arXiv hep-th/0603082
\bibitem{hawking}Hawking, S. W., Commun. Math. Phys. 43 199 (1975). 
\bibitem{adler1}Adler, R. J., Chen, P., Santiago, D. I., Gen. Rel. Grav., 34 2101 (2001). 
\bibitem{modak}Modak, S. K., Phys. Rev. D, 90 p.198 (2014). 
\bibitem{torres}Torres, R., Fayos, F., Lorente-Espin, O., Phys. Lett. B, 720 p.198 (2013).  
\bibitem{fabbri}Fabbri, A., Navario-Salas, J.: Modeling Black Hole Evaporation London: Imperial College Press
\bibitem{gingrich}Gingrich, D. M., J. Phys. G, 37 105008 (2010).  
\bibitem{landsberg}Landsberg, G., J. Phys. G, 32R 337 (2006).
\end{thebibliography}

\section*{Acknowledgment}
M. J. Soleimani would like to give an special thanks to Dr. E. C. Vagenas for his useful discussion. The paper is supported by University of Malaya (Grant No. Ru-023-2014) and University Kebangsaan Malaysia (Grant No. FRGS/2/2013/ST02/UKM/02/2 coordinated by Dr. Geri Kibe Gopir).


\end{document}